\documentstyle[aps,twocolumn]{revtex}
\begin{document}
\draft

\maketitle
\narrowtext

When a quasi-particle Andreev reflects from a normal-superconducting
(N-S) interface, the phase of the outgoing excitation is shifted by
the phase of the superconducting order parameter [1]. Consequently
if a phase coherent normal conductor is in contact with two
superconductors with order parameter phases $\phi_1$, $\phi_2$,
transport properties will be oscillatory functions of the phase
difference $\phi=\phi_1-\phi_2$.
Spivak and Khmel'nitskii [2] argued that the ensemble average
of the electrical conductance
of a diffusive structure would have the form
$\langle G(\phi) \rangle=A+B\cos(2\phi)$
were $A$ and $B$ are positive constants of magnitude less than the
quantum of conductance $2e^2/h$.
Thus the ensemble averaged conductance is predicted to have
a {\bf maximum} at zero phase and to have a fundamental periodicity of $\pi$.
Later it was noted [3] that the conductance $G$ of a single sample
should have a periodicity of $2\pi$,
with an amplitude $<2e^2/h$, although no comment on the possibility
of a zero phase minimum was offered.
The theories of references [2,3] were formulated before exact formulae [4,5]
for the electrical conductance of N-S-N structures were available and
therefore,
to avoid significant supercurrents arising from Andreev scattering, were
restricted to the regime
where the distance between the superconductors is greater than
the thermal coherence length $L_T=\sqrt{\hbar D/k_BT}$.

The aim of this Letter is to address two crucial questions
raised by recent experiments on such interferometers [6-10].
The first concerns the nature of the zero phase extremum in
$G(\phi)$, which in  the theory of [2] and all
experiments using normal contacts
is  found to be a maximum,
whereas in the low temperature limit $M<L_T$, theory [11]
 predicts that minima are allowed. A second
concerns the amplitude of oscillation, which in the
experiments [7,8,10] is
found to be smaller than or of order  $2e^2/h$, but in those of
references [6,9] is several multiples of $2e^2/h$. If the latter
are not an artifact
of the experiments, then superconductivity enhanced quasi-particle
interference effects (SEQUINs) of this kind
should be present in the low temperature theory of references [4,5,11].
 In what follows we
demonstrate that large amplitude SEQUINs are indeed obtainable from exact
solutions of the Bogoliubov - de Gennes equation and  highlight
conditions under which giant oscillations should be observable.
Remarkably we predict that the amplitude vanishes for both very dirty
and very clean samples and that to obtain a large effect,
a degree of normal scattering must be introduced, in order that
an approximate sum rule is broken. Further we predict that the
experiments carried out to date are far from optimal and that
 amplitudes of oscillation many orders
of magnitude greater than $e^2/h$ are attainable.

For simplicity we consider the zero temperature limit,
where the electrical conductance between two normal reservoirs
can be written [4,5] (in units of $2e^2/h$),
$$G=T_0+T_a + {{2(R_a R'_a -T_a T'_a)}\over {R_a+R'_a+T_a+T'_a}}
\eqno{(1).}$$
In this expression,
 $R_0, T_0$
($R_a, T_a$) are the coefficients
for normal (Andreev)
reflection and transmission for zero energy
quasi-particles from reservoir $1$, while
$R'_0, T'_0$ ($R'_a, T'_a$) are corresponding coefficients for
quasi-particles from reservoir $2$. If each of the external leads connecting
the
reservoirs to the scatterer  contains $N$ open channels, these satisfy
$R_0+T_0+R_a+T_a=R'_0+T'_0+R'_a+T'_a=N$
and
$T_0+T_a=T'_0+T'_a.$ Furthermore, in the absence of a magnetic field,
all reflection
coefficients are even functions of $\phi$, while the transmission
coefficients satisfy $T'_0(\phi)=T_0(-\phi)$,
$T'_a(\phi)=T_a(-\phi)$. Consequently on quite general grounds,
in the absence of a field, $G$ is predicted to be an even function of
$\phi$ [12].

Figure 1 shows three examples of
 interferometers, for which results are presented
below. Each has two superconducting regions with definite phases $\phi_1$
and $\phi_2$, in contact with a normal region (shown shaded).
In each case the scattering region is connected to normal, external
current carrying leads, with $N$ conducting channels. In figures
1a and 1b, a normal barrier (shown black)
is placed at the N-S interface.
In what follows we show numerical results
 obtained
from a tight binding model, with diagonal elements $\epsilon_i$ and
nearest neighbour hopping elements of magnitude unity.
In the external leads $\epsilon_i=0$ and in the single line of sites
forming the barrier $\epsilon_i=\epsilon_b$. In a disordered region
of the sample, $\epsilon_i$ is chosen to be a random number uniformly
distributed over the interval $\pm W$.
As discussed in reference [5], the conductance is obtained
by first evaluating the quantum mechanical scattering matrix
and then evaluating the zero temperature conductance formula (1).
The transfer matrix codes used [5] are extremely versatile and can
be used to analyze arbitrary geometries, with multiple contacts.

For the structure of figure 1a, figure 2 shows numerical
results for giant oscillations in
the electrical conductance $G(\phi)$.
In the absence of a barrier $(\epsilon_b=0)$, the
amplitude of oscillation (in units of $2e^2/h$) is negligible
compared with unity, whereas for
$\epsilon_b=1,\, 2,\, 3$ a large amplitude oscillation is present.
As the barrier strength
$\epsilon_b$
increases, the amplitude of oscillation initially increases to a value
of order $Ne^2/h$, before decreasing in proportion to
the zero phase conductance.
 These results show that at intermediate barrier strengths,
the relative  amplitude as well as the absolute amplitude is optimised.
Figure 3  shows results for the phase periodic conductance of structures
1b (solid line) and 1c (dashed line). For the structure 1b in
figure 3 we have presented results for the most favourable
barrier strength, in the absence of disorder. Introducing normal disorder
or changing the barrier strength decreases the amplitude of oscillation.
In the case of structure 1c
in figure 3 there is no barrier, but a disorder comparable to the
experiments of reference [9] has been used. We have examined the structures
in figures 1b and 1c for a variety of disorders and barrier strengths and
in no case have found an amplitude which is more than a few percent of
$N 2e^2/h$.

The crucial role of normal scattering in
optimising this effect can be understood through
 a multiple scattering description of a N-S
interface, which captures the essential physics of
interferometers.
Consider first an idealisation of the structure of
figure 1a, in which the distance $M$ between the superconductors vanishes
and therefore for a long enough sample there is no quasi-particle
transmission. In this limit the total resistance reduces to a sum of two
measureable boundary resistances and in what follows, we therefore
focus attention on the left-hand boundary conductance [13]
 $$G_{\rm B}(\phi)=2R_a=2{\rm Tr}\,\, r_ar_a^\dagger=
\sum_{i,j=1}^N(R_a)_{ij}\eqno{(2)},$$
where $(R_a)_{ij}=\vert(r_a)_{ij}\vert^2$ is the Andreev reflection
probabiltity from channel $j$ to channel $i$.
As in equation (1), the Andreev reflection coefficient is of the form
$R_a=R_{\rm diag}+R_{\rm off-diag}$
where
$R_{\rm diag}=\sum_{i=1}^N (R_a)_{ii}$ and
$R_{\rm off-diag}$ is the remaining contribution from
inter-channel scattering,
$R_{\rm off-diag}=\sum_{i\ne j=1}^N (R_a)_{ij}$.

In the absence of disorder, for $M=0$ and $\phi=0$, translational
symmetry in the
direction transverse to the current flow guarantees that
$R_{\rm off-diag}=0$. For the system of figure 1a, with no barrier,
 no disorder and $M=45$,
figure 4b, shows the behaviour of the coefficients
 $(R_a)_{ij}$ for $i\ne j$ and demonstrates that even for finite $M$,
off-diagonal
scattering at $\phi=0$ is negligible. This figure
  leads us to a second  observation,
 namely that even for non-zero $\phi$, almost all of the
 off-diagonal coefficients are negligibly small and that a given
 channel $i$ couples strongly to at most one other channel $j$.
Consequently in the absence of disorder,
a multiple scattering description involving pairs of channels captures the
essential physics.

Consider a normal barrier to the left of
a N-S interface.
Particles (holes) impinging on the normal scatterer are described by
a scattering matrix $s_{pp}$,
$(s_{hh})$,
and those arriving at the N-S interface by a reflection matrix $\rho$,
where
$$
s_{pp}=\left(
\matrix{
r_{pp} &t'_{pp}\cr
t_{pp} &r'_{pp}\cr
}\right)
,~~~~
\rho=\left(
\matrix{
\rho_{pp} &\rho_{ph}\cr
\rho_{hp} &\rho_{hh}\cr
}\right).$$
The elements of $s$ and $\rho$ are themselves  matrices  describing
scattering between open
channels  of the external leads. For an ideal interface,
where Andreev's approximation is valid [1], $\rho_{pp}$ and
$\rho_{hh}$
are negligible and in what follows will be set to zero.
As a consequence, $\rho_{hp}$ and $\rho_{ph}$ are unitary
and
one obtains [14]
$
r_a=t'_{hh} \rho_{hp} M_{pp}^{-1}t_{pp}
,$
with
$
M_{pp}=1-r'_{pp} \rho_{ph} r'_{hh} \rho_{hp}.$
In contrast with the analysis of [14], where $\rho_{hp}$ is proportional
to the unit matrix, the interference
effect of interest here is contained in the fact that $\rho_{hp}$
induces off-diagonal scattering.
Substituting $r_a$ into equation (2) and taking advantage of
particle-hole symmetry at $E=0$, yields
$$
G=2 Tr\left(T Q^{-1} T (Q^{\dagger})^{-1}\right)    \eqno{(3)}
$$
where
$
Q=\rho_{ph}^t+(r')_{pp}\rho_{ph}(r')_{pp}^{\dagger},$
with $T=t_{pp}t_{pp}^{\dagger}$ the transmission matrix of the
normal scattering region. This multiple scattering formula for the
boundary conductance is valid in the presence of an arbitrary number of
channels and in any dimension. Notice that
if $T$ is equal to the unit matrix,
 $Q=\rho_{ph}^t$ and therefore
$G=2 N$,
irrespective of the phase periodic nature of $\rho_{ph}$.
This demonstrates that at a clean interface, whatever the phase,
the approximate unitarity of $\rho_{ph}$ yields the sum rule
$R_{\rm diag}+R_{\rm off-diag}=N$ and therefore
the conductance is independent
of $\phi$.
More generally, whenever normal reflection ($R_0$)
 and Andreev transmission ($T_a$) are negligibly small,
 unitarity imposes the sum rule
$T_0+R_a=N$ and since in this limit equation (1)
reduces to
$G=T_0+R_a$, the amplitude of oscillation must vanish.

Equation (3) is very general and makes no assumption
 about the nature of matrices $\rho_{ph}$ and $s_{pp}$.
We now introduce a  two-channel model in which
 $\rho_{ph}$  is chosen to be an arbitrary two dimensional unitary
matrix.
In the absence of disorder,  $t_{pp}$ and $r_{pp}$ are
 diagonal and therefore
the only interchannel coupling is provided by $\rho_{ph}$.
 Substituting these
matrices into equation (3), yields an expression for
$r_a$ involving a single phase $\theta$, whose value is a linear
combination of
  phase shifts due to normal reflection at the barrier, Andreev
reflection at the N-S interface and the phase
accumulated by an excitation travelling from the barrier to
the interface. In what follows, we present results for the average
value of ${(R_a)}_{ij}$, obtained by integrating
over $\theta$. For a given value of $\phi$, once the normal
barrier transmission
coefficients $T_1$ and $T_2$ of the two channels are chosen,
$(R_a)_{ij}$ is completely determined.

For the structure of figure 1a,
figures 4a and 4b show numerical results
for the diagonal  $(R_a)_{ii}$ and off-diagonal coefficients
$(R_a)_{ij}$ $(i\ne j)$ respectively.  Notice that at zero phase,
most of the  diagonal coefficients $(R_a)_{ii}$ are close to unity,
although a small number of order $ N \vert\Delta\vert/E_F$, where
$\vert\Delta\vert$ is the order parameter magnitude and $E_F$ the
Fermi energy, are suppressed, due to a breakdown of Andreev's
approximation for low angle scattering [15].
This slight breakdown of Andreev's approximation yields a small
amplitude oscillation even in the absence of normal potential scattering, but
as emphasized by figure 2, the fractional amplitude is negligible.
Figures 5a and 5b show corresponding results for the diagonal and
off-diagonal coefficients in the presence of a barrier. At $\phi=0$,
there is no coupling between the channels and the scattering properties
are those of $N$ independent channels, each with a barrier
transmission coefficient $T_i$. The spectrum of the coeffcients
 depends in detail on the shape of the barrier.
The inset of figure 5a shows the boundary conductance $G(\phi )$
obtained by summing the curves in figures 5a nad 5b.
 Figures 5c and 5d
show results from the two channel evaluation of equation (3),
obtained by choosing 10 pairs of transmission coefficients $T_1$,
$T_2$, with $T_2=0.2T_1$. Clearly the qualitative features of the exact
simulation are reproduced by this simple two channel analysis.

As shown in figure 2, in the absence of zero phase
inter-channel scattering, the zero
phase extremum is a minimum. From figure 4, it is clear that the
nature of the extremum is the result of a competition between
diagonal Andreev reflection coefficients, which exhibit a zero phase
 maximum and
off-diagonal coefficients which possess a zero phase minimum.
Analytically we find that,
in the absence of disorder,
 the second derivative of two channel conductance
satisfies $d^2G/d\phi^2 \ge 0$, for all barrier strengths.
In contrast, figure 6 shows numerical results for the structure of figure
1a, with $M=45$, $M'=50$, $M''=15$, but with the barrier replaced by
 a disordered normal square of width 30 sites.
This shows that replacing the barrier by a disordered region, causes
the extremum of the
off-diagonal coefficients to switch from a
minimum to a maximum. In addition, the channels no longer couple in pairs
and a complete multi-channel scattering description is needed.
It is noted [3,5] that changing the microscopic impurity configuration
shifts $G(\phi )$ by an amount of order $e^2/h$, so in the presence of giant
oscillations, the nature of the zero phase extremum is insensitive
to such changes.

In summary, we have demonstrated that giant SEQUINs are obtainable from
exact solutions of the Bogoliubov - de Gennes equation and can
be observed by breaking
a crucial sum rule.
Remarkably, the structure of figure 1c used in
the experiments of [9] shows only a small percentage effect, which
nevertheless, in view of the large number of channels in these
experiments, yielded an amplitude greater than $2e^2/h$. We predict
that a more optimal structure is obtained in the presence
of a normal barrier at the interface and  that in metallic samples,
with very large $N$, the amplitude of oscillation could be orders of
magnitude larger than $2e^2/h$.

Since carrying out this work, we became aware of a related paper on
 diffusive conductors [16], where it is predicted that $(R_a )_{ii}
 \gg (R_a )_{ij}$ ($i\neq j$). This phenomenon of giant diagonal scattering
(i.e. backscattering ) was offered as the origin of large oscillations
in $G ( \phi )$. In this Letter, we predict (see eg. fig.4) that giant
oscillations can occur even when diagonal and off-diagonal
 scattering coeffiecients are comparable.

This work is supported by the EPSRC, the EC Human Capital and Mobility
Programme, NATO, the MOD and the Institute
for Scientific Interchange (Torino).
It has benefited from useful conversations with V. Petrashov,
 F. Sols, and C.W.J. Beenakker.



\begin{figure}
\caption{
Three possible interferometers, each with
 two superconducting regions of width $M''$. In figures 1a and 1b, the
 superconductors are separated by a distance $M$ and in 1c by a distance
 $3M$.
In figure 1a, the scattering region is connected to normal, external
current carrying leads, of width $M +2M''$, in figures 1b and 1c of width
$M$. In figures
1a and 1b, a normal barrier (shown black)
is placed at the N-S interface. The current flows from left to right
between external reservoirs with potentials $\mu_1$ and $\mu_2$.
In the tight binding model used in the numerical simulations,
the barrier comprises a
line of sites with diagonal elements $\epsilon_i=\epsilon_b$.
}
\end{figure}

\begin{figure}
\caption{
Numerical results for the conductance $G$ of the structure of figure 1a,
with $M=45$, $M'=50, M''=15$, and number of open channels $N=45$.
Results are shown for  barrier potentials
$\epsilon_b=0,\, 1,\, 2,\, 3$. The number adjacent a given curve is the
correponding value of $\epsilon_b$.
}
\end{figure}

\begin{figure}
\caption{
Figure 3 shows results for the conductance $G$ of the structure of figure
 1b (solid line: left scale)
with $M=M'= M''=10$, $W=0$, and 1c (dashed line: right scale)
with $M=M'= M''=10$, $W=0$. $N=10$ in both cases.
For the structure 1b results are shown
 for a barrier potential $\epsilon_b=2$.
}
\end{figure}

\begin{figure}
\caption{
Figure 4a shows
numerical results for the diagonal Andreev reflection coefficients
$(R_a)_{ii}$ of the structure of figure 1a,
with $M=45$, $M'=50$, $M''=15$, $N=45$, and no barrier present,
($\epsilon_b=0$.)
Figure 4b shows corresponding results for the off-diagonal coefficients
$(R_a)_{ij}$ with $i\ne j$.
}
\end{figure}

\begin{figure}
\caption{
Figure 5a shows
numerical results for the diagonal Andreev reflection coefficients
$(R_a)_{ii}$ of the structure of figure 1a,
with  $M=45$,$M'=50$, $M''=15$, $N=45$, and barrier potential,
$\epsilon_b=2$.
Figure 5b shows corresponding results for the off-diagonal coefficients
$(R_a)_{ij}$ with $i\ne j$.
Figures 5c and 5d show analytical results from a two channel calculation.
}
\end{figure}

\begin{figure}
\caption{
Numerical results obtained from a tight binding model
of the structure of figure 1a, but with the barrier
replaced by a disordered region of length 30 sites.
In these simulations, $M=45$, $M'=15$, $M''=50$, $N=45$,
$\Delta_0=0.1$, and the disorder is $W=2.8$.
}
\end{figure}

\begin{thebibliography}{99}
\bibitem{12.} It should be noted that the figures of reference [11] also
contain
a small number of plots which are not an even function of the phase. These
are an artifact of the graph plotting code used to construct the curves and
should be ignored. An examination of data files produced by the
numerical simulations in reference [11] show that $G(\phi)$ is always
an even function of $\phi$.
\end{thebibliography}
\end{document}